\documentclass[final,3p,times,sort&compress]{elsarticle}
\usepackage{multirow,setspace,times,amssymb,amsmath,graphicx,color,rotating,subfigure,url}
\usepackage{lineno}
\usepackage{natbib}
\usepackage{amsmath}
\usepackage[bookmarks=true,colorlinks,linkcolor=red,anchorcolor=blue,citecolor=green,unicode]{hyperref}
\usepackage{bookmark}
\usepackage[normalem]{ulem}
\usepackage{epstopdf}
\vfuzz=\maxdimen
\journal{Entropy}

\begin{document}

\begin{frontmatter}

\title{Information transfer between stock market sectors: A comparison between the USA and China}
\author[BS]{Peng Yue}
\author[AU,SUFE]{Yaodong Fan}
\author[MMU]{Jonathan A. Batten}
\author[BS,SS,RCE]{Wei-Xing Zhou\corref{cor1}}
\cortext[cor1]{Corresponding author. Address: 130 Meilong Road, P.O. Box 114,
              East China University of Science and Technology, Shanghai 200237, China.}
\ead{wxzhou@ecust.edu.cn} %

\address[BS]{School of Business, East China University of Science and Technology, Shanghai 200237, China}
\address[AU]{School of Business, University of Technology Sydney, Sydney NSW 2007, Australia}
\address[SUFE]{College of Business, Shanghai University of Finance and Economics, Shanghai 200433, China}
\address[MMU]{School of Economics, Finance and Banking, College of Business, Universiti Utara Malaysia, 06010 UUM Sintok, Kedah, Malaysia}
\address[SS]{Department of Mathematics, East China University of Science and Technology, Shanghai 200237, China}
\address[RCE]{Research Center for Econophysics, East China University of Science and Technology, Shanghai 200237, China}

\begin{abstract}
Information diffusion within financial markets plays a crucial role in the process of price formation and the propagation of sentiment and risk. We perform a comparative analysis of information transfer between industry sectors of the Chinese and the USA stock markets, using daily sector indices for the period from 2000 to 2017. The information flow from one sector to another is measured by the transfer entropy of the daily returns of the two sector indices. We find that the most active sector in information exchange (i.e., the largest total information inflow and outflow) is the {\textit{non-bank financial}} sector in the Chinese market and the {\textit{technology}} sector in the USA market. This is consistent with the role of the non-bank sector in corporate financing in China and the impact of technological innovation in the USA. In each market, the most active sector is also the largest information sink that has the largest information inflow (i.e., inflow minus outflow). In contrast, we identify that the main information source is the {\textit{bank}} sector in the Chinese market and the {\textit{energy}} sector in the USA market. In the case of China, this is due to the importance of net bank lending as a signal of corporate activity and the role of energy pricing in affecting corporate profitability. There are sectors such as the {\textit{real estate}} sector that could be an information sink in one market but an information source in the other, showing the complex behavior of different markets. Overall, these findings show that stock markets are more synchronized, or ordered, during periods of turmoil than during periods of stability.
\end{abstract}


\end{frontmatter}

\section{Introduction}
\label{S1:Intro}

Complex systems, such as financial markets, are usually composed of many subsystems; in the case of financial markets, information flows and interactions within the market itself are rarely investigated even though they are critical to driving the complex dynamics of the complex system as a whole. Many methods have been proposed to unveil these different relationships among subsystems, such as correlations including simple correlation analysis \citep{Mantegna-Stanley-2000,Zhang-Zhang-Shen-Zhang-2018-Complexity}, Granger causality \citep{Shan-Pappas-2000-AFE}, nonparametric approaches such as the thermal optimal path method \citep{Sornette-Zhou-2005-QF,Meng-Xu-Zhou-Sornette-2017-QF,Xu-Zhou-Sornette-2017-JIFMIM}, and mutual information analysis \citep{Dionisio-Menezes-Mendes-2004-PA,Abigail-2013-PA,Fiedor-2014-PRE}. These different approaches have their own advantages and limitations. Importantly, while Granger causality is commonly used to identify time-varying single or bidirectional causality in economics, it is sensitive to sample period selection and complexity in the underlying time series, as well as having other issues \citep{Ghysels-Hill-Motegi-2016-JE,Gotz-Hecq-Smeekes-2016-JE}. 

In this paper, we use an alternative approach termed transfer entropy to identify the information transfers between industrial sectors in the world's two largest economies: the USA and China. Transfer entropy, as a kind of log-likelihood ratio \citep{Barnett-Bossomaier-2012-PRL}, is a measure that quantifies information flow based on the probability density function (PDF). {{Better than correlations or Granger causality, transfer entropy}} not only identifies the direction of the information flow but also quantifies the flows between different subsystems. {{In other words, it is capable of quantifying the strength and direction of the interaction between different subsystems at the same time.}} This approach has found wide application \citep{Schreiber-2000-PRL, Ai-2014-Entropy, Hu-Zhao-Ai-2016-Entropy, Yook-Chae-Kim-Kim-2016-PA, BorgeHolthoefer-Peera-Goncalves-GonzalezBailon-Arenas-Moreno-Vespignani-2016-SciAdv,Zhang-Lin-Shang-2017-FNL, Toriumi-Komura-2018-CE, Servadio-Convertino-2018-SciAdv,Zhang-Shang-Xiong-Xia-2018-FNL}. Furthermore, variation and extensions of transfer entropy have been developed that are suitable for different situations \citep{He-Shang-2017-PA}, such as symbolic transfer entropy \citep{Staniek-Lehnertz-2008-PRL}.

There are many studies adopting the concept of transfer entropy to economic systems such as financial time series
\citep{Marschinski-Kantz-2002-EPJB,Mao-Shang-2017-CNSNS,Zhang-Lin-Shang-2017-FNL}, stock market indices \citep{Dimpfl-Peter-2013-SNDE, Kwon-Yang-2008-EPL}, composite index and the constituent stocks~
 \citep{Kwon-Yang-2008-PA,Kwon-Oh-2012-EPL}, and indices of industry sectors of a stock market \citep{Oh-Oh-Kim-Kwon-2014-JKPS}.

Stock price fluctuations reflect both global and local news as well as news within a subsystem. There are also well-known calendar anomalies related to business cycle and market participants sector rotations \citep{Leibon-Pauls-Rockmore-Savell-2008-PNAS}. In a related work, Oh et al. investigated the information flows among different sectors of the Korean stock market \citep{Oh-Oh-Kim-Kwon-2014-JKPS}. They measured the amount of information flow and the degree of information flow asymmetry between industry sectors around the subprime crisis and identified the insurance sector as the key information source after the crisis. Although the authors do not attribute a economic basis for this finding, it is likely linked to the insurance sector acting as a leading indicator of risk in the economy. In this work, their analysis is extended and a comparative study is performed on the information transfer among different industry sectors of the Chinese and the USA stock markets. These two markets are respectively the largest emerging and developed stock markets associated with the two largest economies in the world.

The rest of this paper is organized as follows. Section \ref{S1:Method:Data} describes the method for calculating symbolic transfer entropy and the sector indices time series for the Chinese and the USA stock markets. Section \ref{S1:Results} presents the empirical results about the information flows between stock market sectors and its relationship with market states. Section \ref{S1:conlude} concludes this work.

\section{Method and Data}
\label{S1:Method:Data}

\subsection{Symbolic Transfer Entropy}
\label{S2:STE}

Schreiber was the first to use transfer entropy to measure information transfer and detect asymmetry in the interactions among subsystems \cite{Schreiber-2000-PRL}. He treated a sleeping human's breath rate time series and heart rate time series as two subsystems and found that the information flow from the heart to the breath signal is dominant. To explore the transfer entropy between two time series, there are various approaches in the literature. We need to briefly summarize what the other approaches are and why the symbolic transfer method is used. We use the symbolic transfer entropy introduced by Staniek and Lehnertz \cite{Staniek-Lehnertz-2008-PRL}. Consider two different daily closing prices time series \{$X_t$\} and \{$Y_t$\}, $t=1,2,\ldots, L$, which have the same length $L$. Closing prices are used to ensure that prices factor in local market news as well as intramarket news from the various sectors. Transfer entropy $T^S_{y\rightarrow{x}}$ assumes that $X_t$ is influenced by the previous $l$ states of the same variable and by the $m$ previous states of variable $Y$, for financial markets, only the day before is important \cite{Sandoval-2014-Entropy}. Hence, we use $l=m=1$ in this study. The procedure to calculate the symbolic transfer entropy $T^S_{y\rightarrow{x}}$ from time series \{$y_t$\} to \{$x_t$\} is briefly described in the following five steps:

First, we adopt the log returns \{${x_t}$\} instead of the original price time series \{$X_t$\} by
\begin{equation}
  x_t\equiv{\ln(X_t)-\ln(X_{t-1})}
\end{equation}
where $X_t$ is the closing price of the index on the $t$th trading day.

Second, the returns are discretized into $q$ nonoverlapping windows of equal length $\Delta$. If there are too many windows, the chance of having particular combinations drops very quickly, making the calculation of probabilities slower and less informative \citep{Sandoval-2014-Entropy}. Hence, it is irrational if $q$ is too large or too small. Marschinski and Kantz consider $q=2$ and $q=3$ in their research \citep{Marschinski-Kantz-2002-EPJB}; Sandoval uses $q=24$ and $q=6$ \citep{Sandoval-2014-Entropy}. We aim at finding the optimal $q$ to maximize the transfer entropy difference between two time series meanwhile minimizing the calculation cost. In our comparative investigations, the parameter $q$ varies from 2 to 22 with a moving step of 1. We find that when $q\ge{15}$, the difference becomes significantly nonzero. Considering the calculation cost and the strength of transfer entropy, in this work, we use $q=15$. We obtain the maximum value $x_{\max}$ and minimum value $x_{\min}$ of the time series $x_t$ under investigation.  The length of each interval is $\Delta_x=[x_{\max}-x_{\min}]/q$ and the $k$th interval is $[x_{\min}+(k-1)\Delta_x, x_{\min}+k\Delta_x)$. Similarly, we repeat the procedure for $y_t$ and its $\Delta=\Delta_y$ is usually different from $\Delta_x$.

Third, the log return time series $\hat{x}$ and $\hat{y}$ are described as
\begin{equation}
  \hat{x}_t=f(x_t)=k_x~~{\mathrm{and}}~~ \hat{y}_t=f(y_t)=k_y, ~~~k_x,k_y=1,2,\cdots,q,
\end{equation}
where $x_t\in{[x_{\min}+(k_x-1) \Delta_x, x_{\min}+k_x \Delta_x)}$ and $y_t\in{[y_{\min}+(k_y-1) \Delta_y, y_{\min}+k_y \Delta_y)}$.

Fourth, the number of elements in the $q$th interval are denoted by $\hat{x}_t^q$ and $\hat{y}_t^q$, 
 respectively, and then calculate the probabilities
$p(\hat{x}_t)=\hat{x}_t^q/(L-1)$ and $p(\hat{y}_t)=\hat{y}_t^q/(L-1)$ and the joint probabilities $p(\hat{x}_t,\hat{y}_t)$, $p(\hat{x}_t,\hat{x}_{t+1})$ and
$p(\hat{x}_{t+1},\hat{x}_t,\hat{y}_t)$.

Fifth, in information theory, different bases of entropy lead to different units of entropy. Base 2 is the most widely applied in transfer entropy for most of empirical works. Therefore, in our study, we use Base 2 to calculate transfer entropy. The symbolic transfer entropy from time series \{${y_t}$\} to time series \{${x_t}$\} is calculated as
\begin{equation}
  T^S_{y\rightarrow{x}}=\sum_{\hat{x}_{t+1},\hat{x}_t,\hat{y}_t}{p(\hat{x}_{t+1},\hat{x}_t,\hat{y}_t)\log_2{\frac{p(\hat{x}_{t+1}|\hat{x}_t,\hat{y}_t)}{p(\hat{x}_{t+1}|\hat{x}_{t})}}},
\label{Eq:TE}
\end{equation}
where the joint probability $p(\hat{x}_{t+1},\hat{x}_t,\hat{y}_t)$ means the probability that the combination of $\hat{x}_{t+1}$, $\hat{x}_t$ and $\hat{y}_t$ occurs, while $p(\hat{x}_{t+1}|\hat{x}_t,\hat{y}_t)$ and $p(\hat{x}_{t+1}|\hat{x}_t)$ are the conditional probabilities that $\hat{x}_{t+1}$ has a particular value when the values of previous samples $\hat{x}_t$ and $\hat{y}_t$ are known and $\hat{x}_t$ is known, respectively.
Since
\begin{equation}
  p(\hat{x}_{t+1}|\hat{x}_t,\hat{y}_t)=\frac{p(\hat{x}_{t+1},\hat{x}_t,\hat{y}_t)}{p(\hat{x}_t,\hat{y}_t)}~~{\rm{and}}~~p(\hat{x}_{t+1}|\hat{x}_t)=\frac{p(\hat{x}_{t+1},\hat{x}_t)}{p(\hat{x}_t)},
\end{equation}
we can simplify Equation~(\ref{Eq:TE}) and obtain
\begin{equation}
  T^S_{y\rightarrow{x}}=\sum_{\hat{x}_{t+1},\hat{x}_t,\hat{y}_t}{p(\hat{x}_{t+1},\hat{x}_t,\hat{y}_t)\log_2{\frac{p(\hat{x}_{t+1},\hat{x}_t,\hat{y}_t)p(\hat{x}_t)}{p(\hat{x}_{t+1},\hat{x}_t)p(\hat{x}_t,\hat{y}_t)}}}.
\end{equation}
This expression is used for the estimation of the symbolic transfer entropy.

\subsection{Data Description}
\label{S2:Data}

To conduct our analysis, we selected two sets of data from two major stock markets: the Chinese and the USA stock market. The Chinese stock
market is the largest emerging market, while the US stock market is the worlds largest developed stock market.

For the Chinese stock market, we retrieved and analyzed the SWS 
 sector indices issued by Shenyin \& Wanguo Securities Co., Ltd. (http://www.swsresearch.com). In total, this gave 28 sector indices of the Chinese stock market, and covered 3508 individual stocks. For each sector index series, there were 4359 daily prices from 4 January 2000 to 29 December 2017. The various sectors with their corresponding six-digit codes include: agriculture and forestry (801010), mining (801020), chemical (801030), steel (801040), non-ferrous metals (801050), electronic (801080), household appliances (801110), food and drink (801120), textile and apparel (801130), light manufacturing (801140), biotechnology (801150), utilities (801160), transportation (801170), real estate (801180), commercial trade (801200), leisure and services (801210), composite (801230), building materials (801710), building and decoration (801720), electrical equipment (801730), national defense (801740), computer (801750), media (801760), communications (801770), bank (801780), non-bank financial (801790), automobile (801880), and mechanical equipment (801890).

For the US stock market, we chose 16 sector indices composed by Thompson Reuters Co., Ltd. (http://www.thomsonreuters.cn). For each sector index time series, there were 4695 daily prices from 3 January 2000 to 29 December 2017. The differences in day count in the two series are due to differences in holidays in the two countries. The stock sectors are appliances resources (M3L), banking/investment services (BIL), cyclical construction producers (YPL), cyclical consumer services (CRL), energy (E2L), food/beverages (FBL), healthcare services (HSL), industrial and commercial services (IVL), industrial goods (IGL), mineral resources (MRL), pharmaceuticals/MD research (PHL), real estate (REL), retailers (RTL), technology (TEL), transportation (TRL), and utilities (U2\$). We use the concept of symbolic transfer entropy to detect and measure the information flows among the return time series of the 28 sector indices of the Chinese stock market and the 16 sector indices of the USA stock market.

\section{Results and Discussion}
\label{S1:Results}
 
\subsection{Symbolic Transfer Entropy and Degree of Asymmetric Information Flow of the Whole Samples}

As mentioned in Section \ref{S1:Intro}, symbolic transfer entropy can proxy for the strength and direction of the information flow between two time series. Following Oh et al. \cite{Oh-Oh-Kim-Kwon-2014-JKPS}, we used the degree of asymmetric information flow to measure the information effect between two stock sectors, which is defined as
\begin{equation}
   \Delta{T}^S_{i\rightarrow{j}}=T^S_{i\rightarrow{j}}-T^S_{j\rightarrow{i}}{\color{red}.}
   \label{Eq:DT}
\end{equation}

\begin{figure}[!h]
  \centering
  \includegraphics[width=6.5cm]{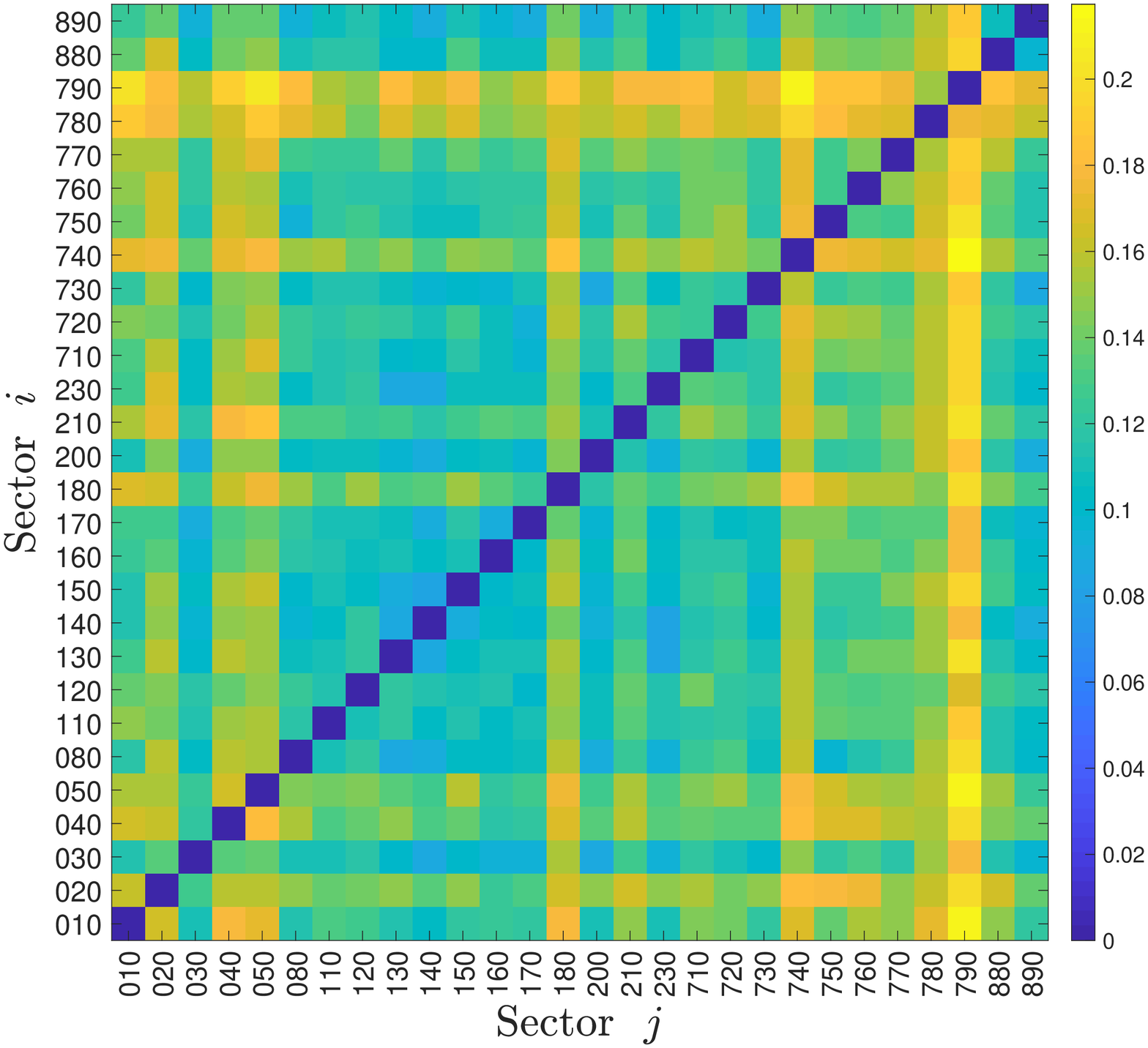} \hskip 0.2cm
  \includegraphics[width=6.5cm]{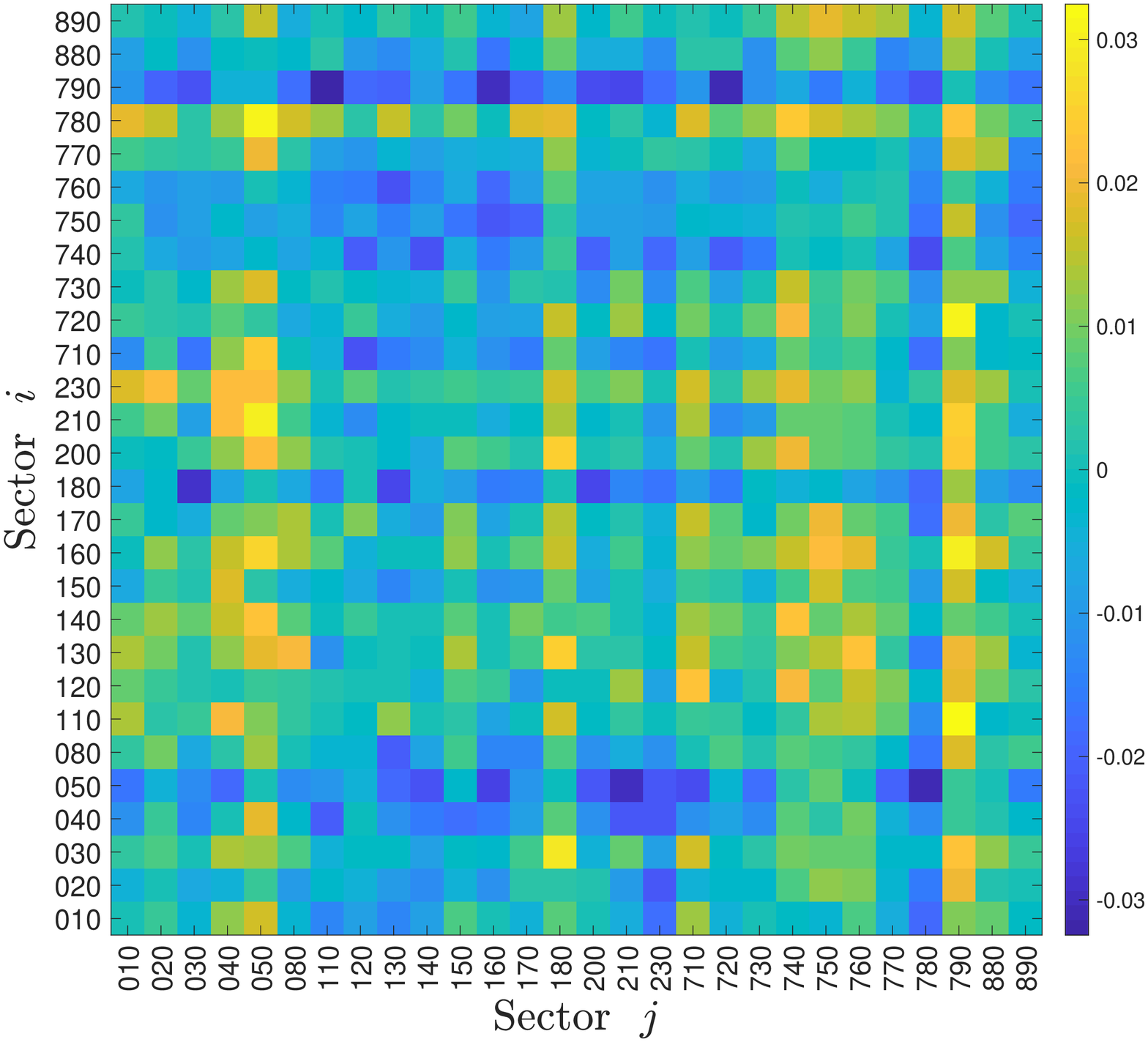}\\
  \includegraphics[width=6.5cm]{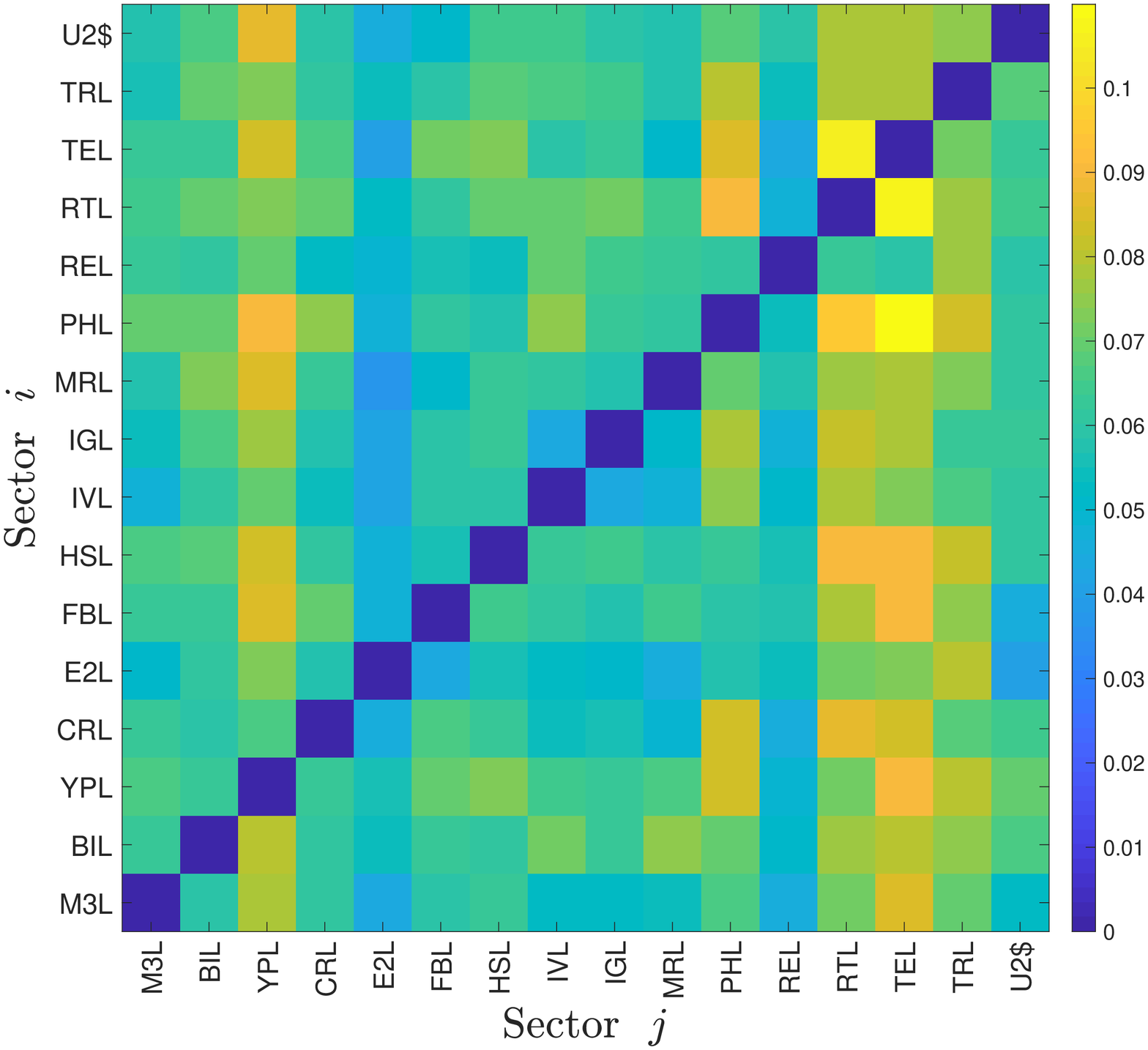} \hskip 0.2cm
  \includegraphics[width=6.5cm]{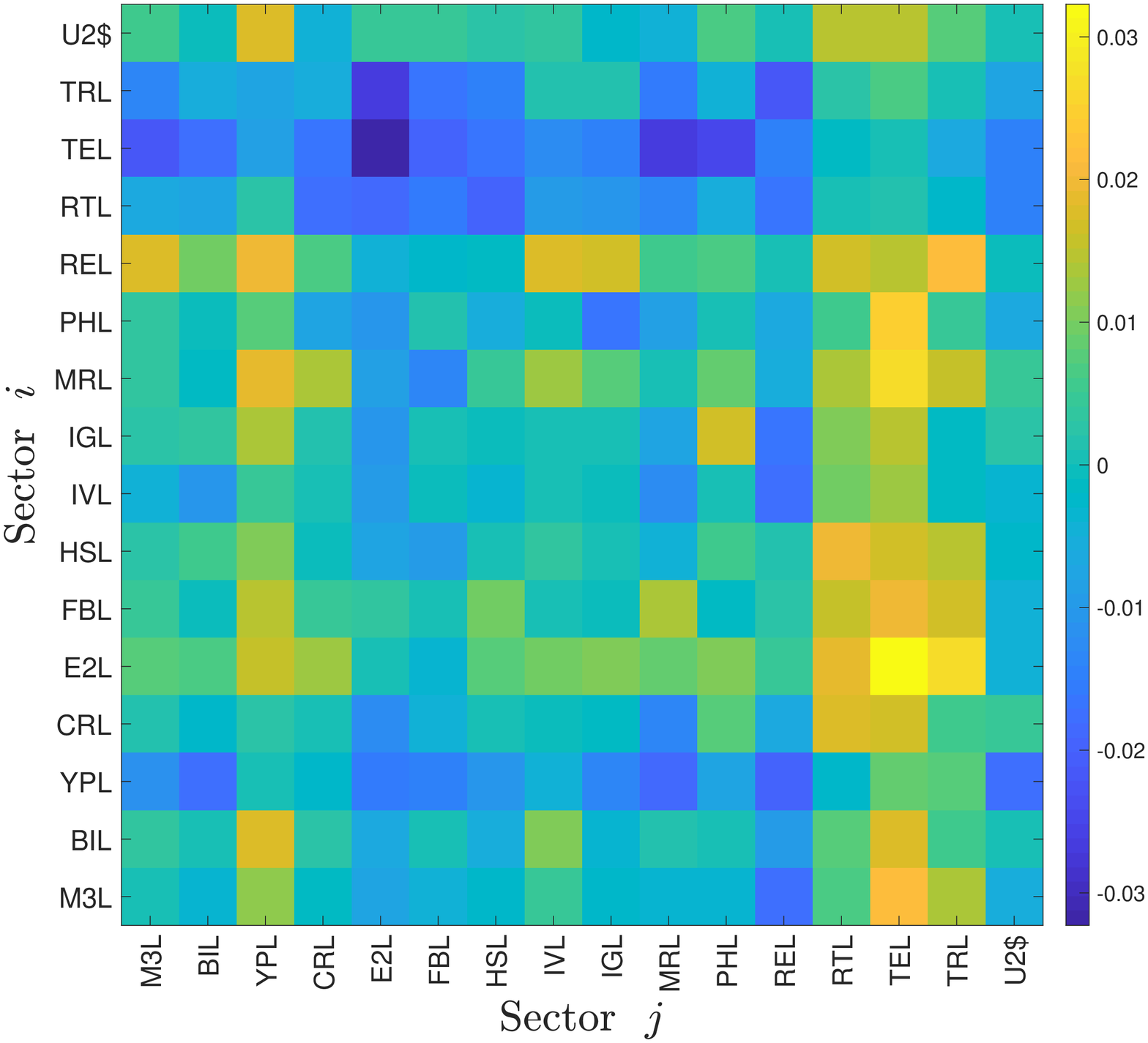}
  \caption{Heat maps of the symbolic transfer entropy matrix $T^S_{i,j}$ (left {matrices}) and the degree of asymmetric information flow $\Delta{T}^S_{i,j}$ (right {matrices}) between the
  28 Chinese stock market sectors (2000-2017, top {matrices}) and the 16 USA stock market sectors (2000-2017, bottom {matrices}). To simplify the label, we use the last three digits
  of each 6-digit code to represent the corresponding Chinese stock market sector.}
  \label{Fig:STE:Sector:Ts:dTs}
\end{figure}

It follows that $\Delta{T}^S_{j\rightarrow{i}}=-\Delta{T}^S_{i\rightarrow{j}}$.
We show the calculation results of our datasets in four heat maps (top row for the Chinese sectors and bottom row for the US sectors) of $T^S_{i,j}$ and $\Delta{T}^S_{i,j}$ in Figure~\ref{Fig:STE:Sector:Ts:dTs}, in which each cell shows the value of $T^S$ (left plot) or $\Delta{T}^S$ (right plot) from sector $i$ to sector $j$. We observe that the values in the diagonal matrices $T^S_{i,j}$ and $\Delta{T}^S_{i,j}$ are zeros, which is trivial and can be understood from the concept of symbolic transfer entropy. We find that the {\textit{non-bank financial}} sector (code 790) has roughly the highest $T^S$ values for both inflows and outflows among the Chinese sectors, and the {\textit{technology}} sector (code TEL) has roughly the highest $T^S$ values for both inflows and outflows among the US sectors; {the {\textit{non-bank financial}} sector comprises three Level 2 sectors in the SWS index system which are {\textit{security}}, {\textit{insurance}}, and {\textit{multivariate financial}}}. This suggests that during the sample period from 2000 to 2017, the {\textit{non-bank financial}} sector and the {\textit{technology}} sector were respectively the most active in the two stock markets. That is, there was more information {exchange between these sectors with the other sectors in their own stock markets than between other sectors in the same stock market}.

\subsection{Average Inflow and Outflow}

For each sector $i$, the average outflow $F_{{\rm{out}},i}$ and inflow $F_{{\rm{in}},i}$ of information can be calculated as follows \citep{Oh-Oh-Kim-Kwon-2014-JKPS}:
\begin{subequations}
\begin{equation}
   F_{{\rm{out}},i}=\frac{1}{n-1}\sum_{p\neq{i}}{T^S_{i\rightarrow{p}}}
\label{Eq:Eq:Fout}
\end{equation}
and
\begin{equation}
   F_{{\rm{in}},i}=\frac{1}{n-1}\sum_{p\neq{i}}{T^S_{p\rightarrow{i}}},
\label{Eq:Fin}
\end{equation}
\label{Eq:Fin:Fout}
\end{subequations}
where the points with $i=j$ are not included.

\begin{figure}[h]
  \centering
  \includegraphics[width=7cm]{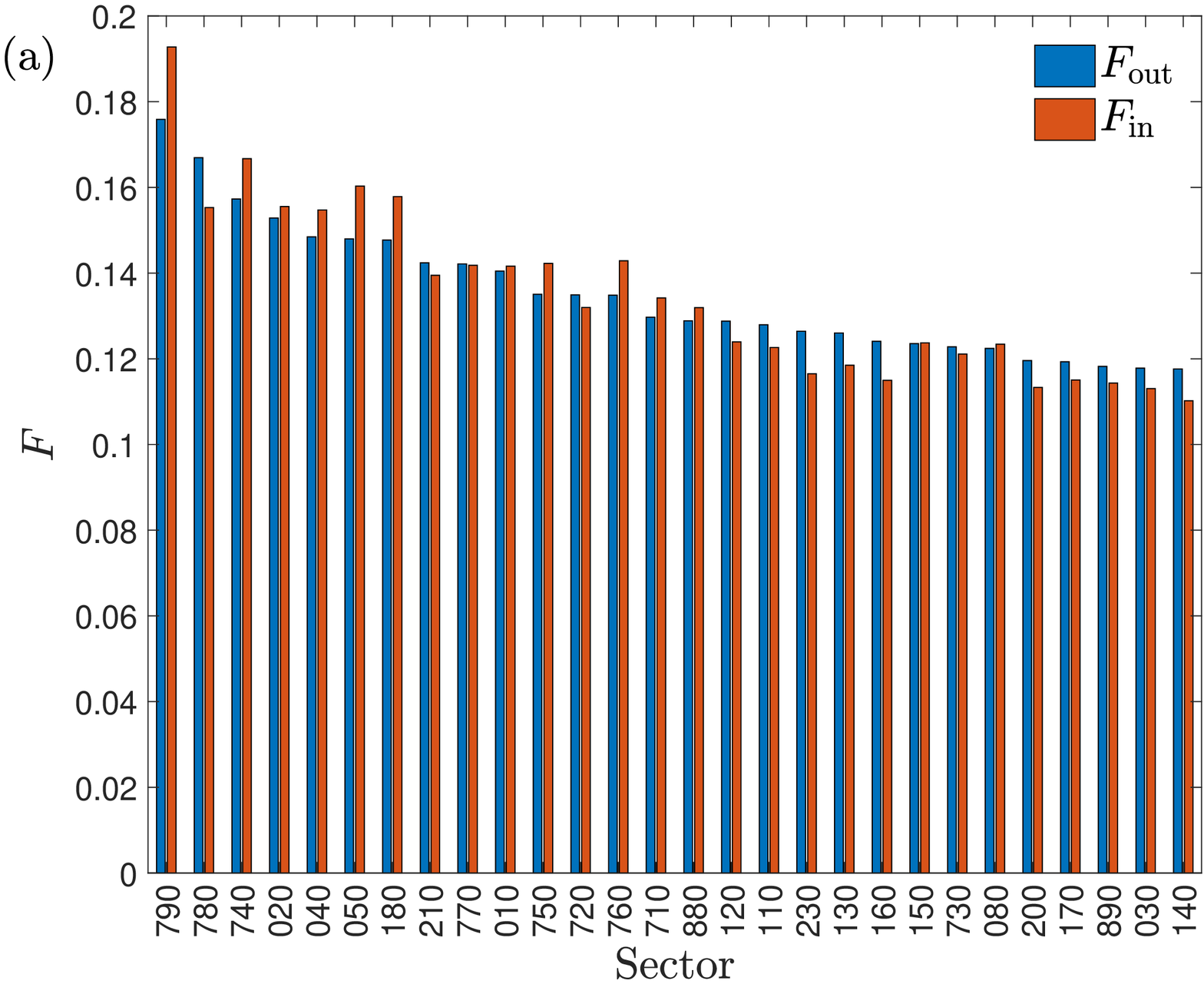}\hskip 0.2cm
  \includegraphics[width=7cm]{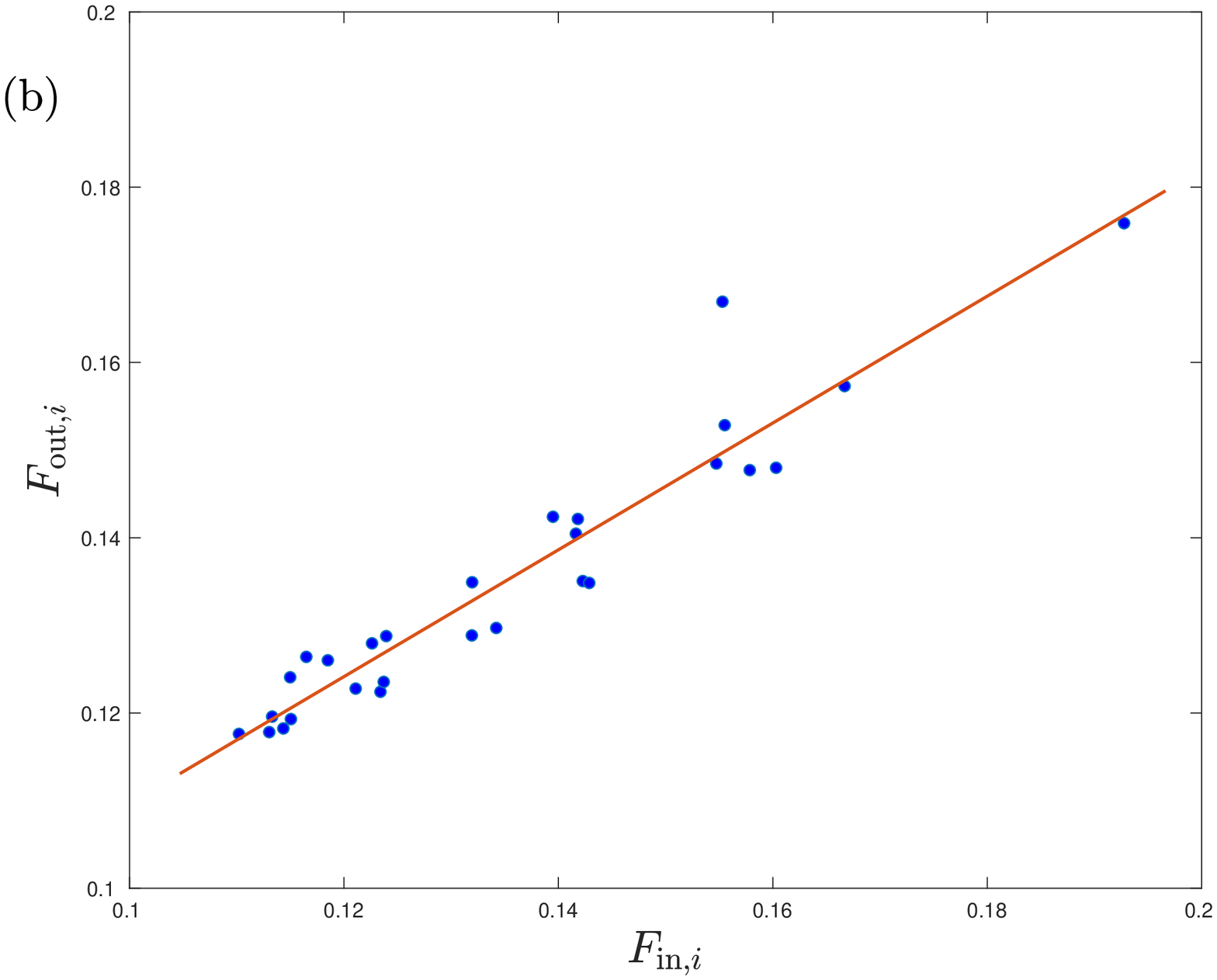}\\
  \includegraphics[width=7cm]{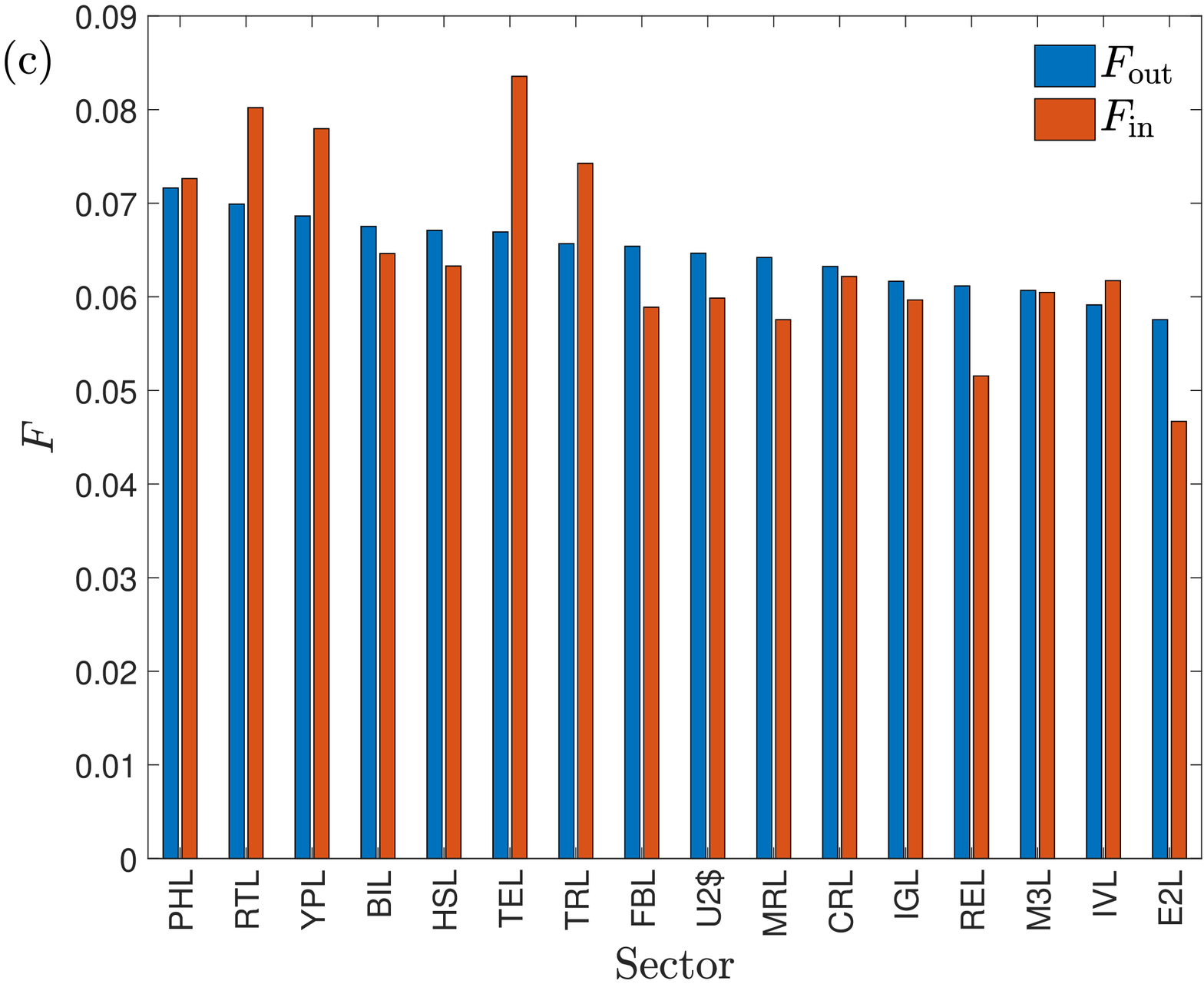}\hskip 0.2cm
  \includegraphics[width=7cm]{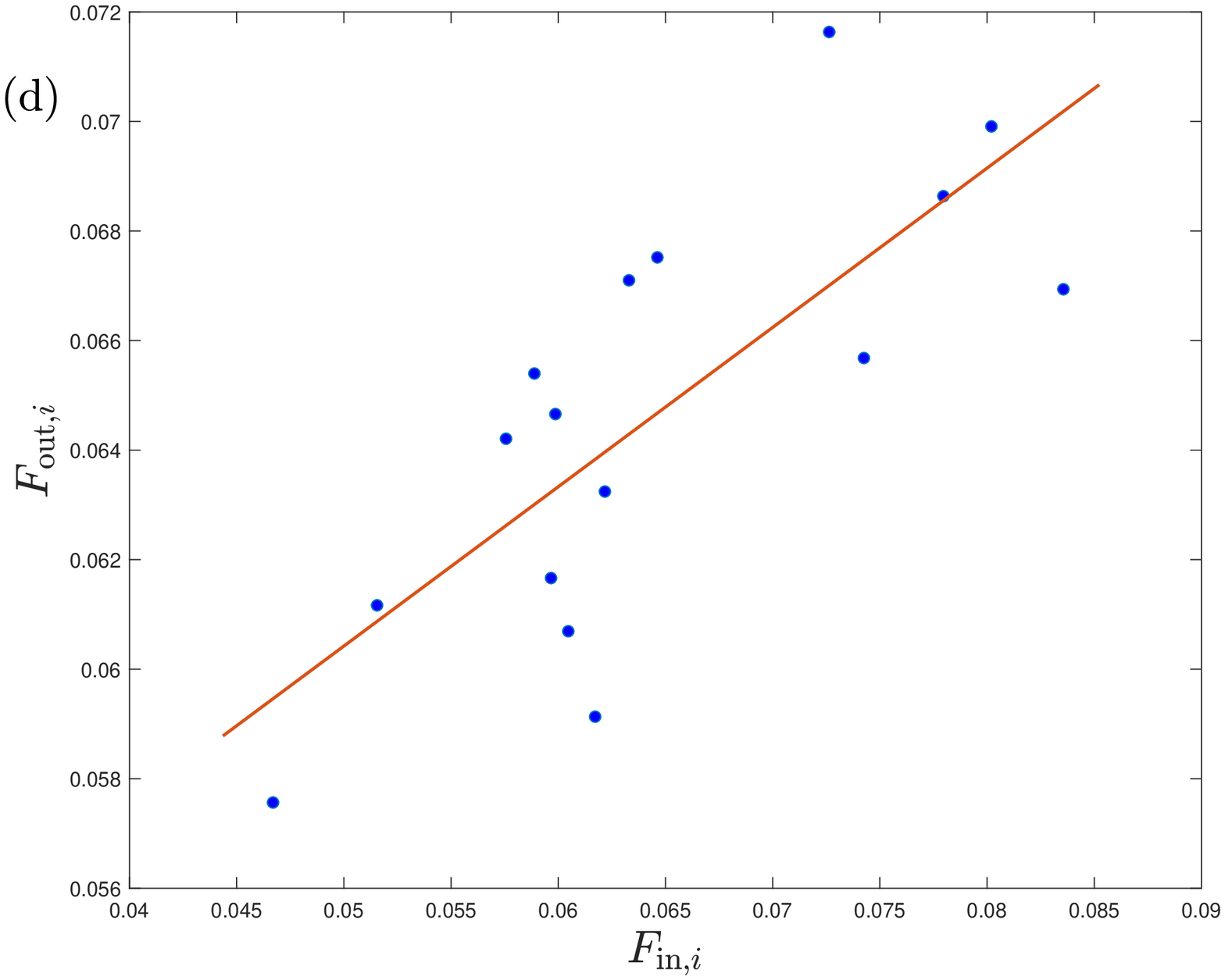}
  \caption{Average outflow $F_{{\rm{out}},i}$ and inflow $F_{{\rm{in}},i}$ of information for stock market sectors. ({\bf a}) Bar chart for the Chinese stock market. ({\bf b}) Scatter
  plot for the Chinese stock market. ({\bf c}) Bar chart for the USA stock market. ({\bf d}) Scatter plot for the USA stock market.}
  \label{Fig:STE:Sector:Fin:Fout}
\end{figure}

Figure~\ref{Fig:STE:Sector:Fin:Fout}a and Figure~\ref{Fig:STE:Sector:Fin:Fout}c show the bar charts of the average information inflows and outflows for all the sectors. This figure confirms that the {\textit{non-bank financial}} sector (code 790) and the {\textit{technology}} sector (code TEL) were the most active sectors in information exchange, respectively. We also find that the more information a sector sends out to other sectors, the more information it receives from others generally. Therefore, the outflow and inflow are positively related to each other. We present in Figure~\ref{Fig:STE:Sector:Fin:Fout}b and  Figure~\ref{Fig:STE:Sector:Fin:Fout}d the scatter plot of $F_{{\rm{out}},i}$ against $F_{{\rm{in}},i}$, which confirms a significant positive correlation. The least-squares regression results in the following linear relationship for the Chinese market are as follows: 
\begin{subequations}
\begin{equation}
   F_{{\rm{out}},i}= 0.724 F_{{\rm{in}},i}+ 0.037,
\label{Eq:Fin:Fout:LinearFit:CN}
\end{equation}
where the $p$-values of the two coefficients are respectively $3\times10^{-15}$ and $2\times10^{-6}$ and the adjusted $R^2$ is 0.908. Similarly, for the USA market we have
\begin{equation}
   F_{{\rm{out}},i}= 0.291 F_{{\rm{in}},i}+ 0.046,
\label{Eq:Fin:Fout:LinearFit:US}
\end{equation}
where the $p$-values of the two coefficients are $6\times10^{-4}$ and $5\times10^{-8}$, respectively, and the adjusted $R^2$ is 0.548. It is clear from this simple estimation that the linear relationship is more significant for the Chinese stock market. {{We argue that the linearity reflects the degree of traders' actions on the idiosyncratic traits of market sectors. The higher linearity of the Chinese stock market implies that the traders in the Chinese market are more irrational, such that their behavior is less reflected in the idiosyncratic traits of market sectors in their decision-making process.}} 
\label{Eq:Fin:Fout:LinearFit}
\end{subequations}

We also use the average degree of asymmetric information flow ${\Delta{F_i}}$ to measure the net information of sector $i$ being sent to other sectors, which is defined as follows
\citep{Oh-Oh-Kim-Kwon-2014-JKPS}:
\begin{equation}
   {\Delta{F_i}}=F_{{\rm{out}},i}-F_{{\rm{in}},i}.
\end{equation}

\begin{figure}[h]
  \centering
  \includegraphics[width=7cm]{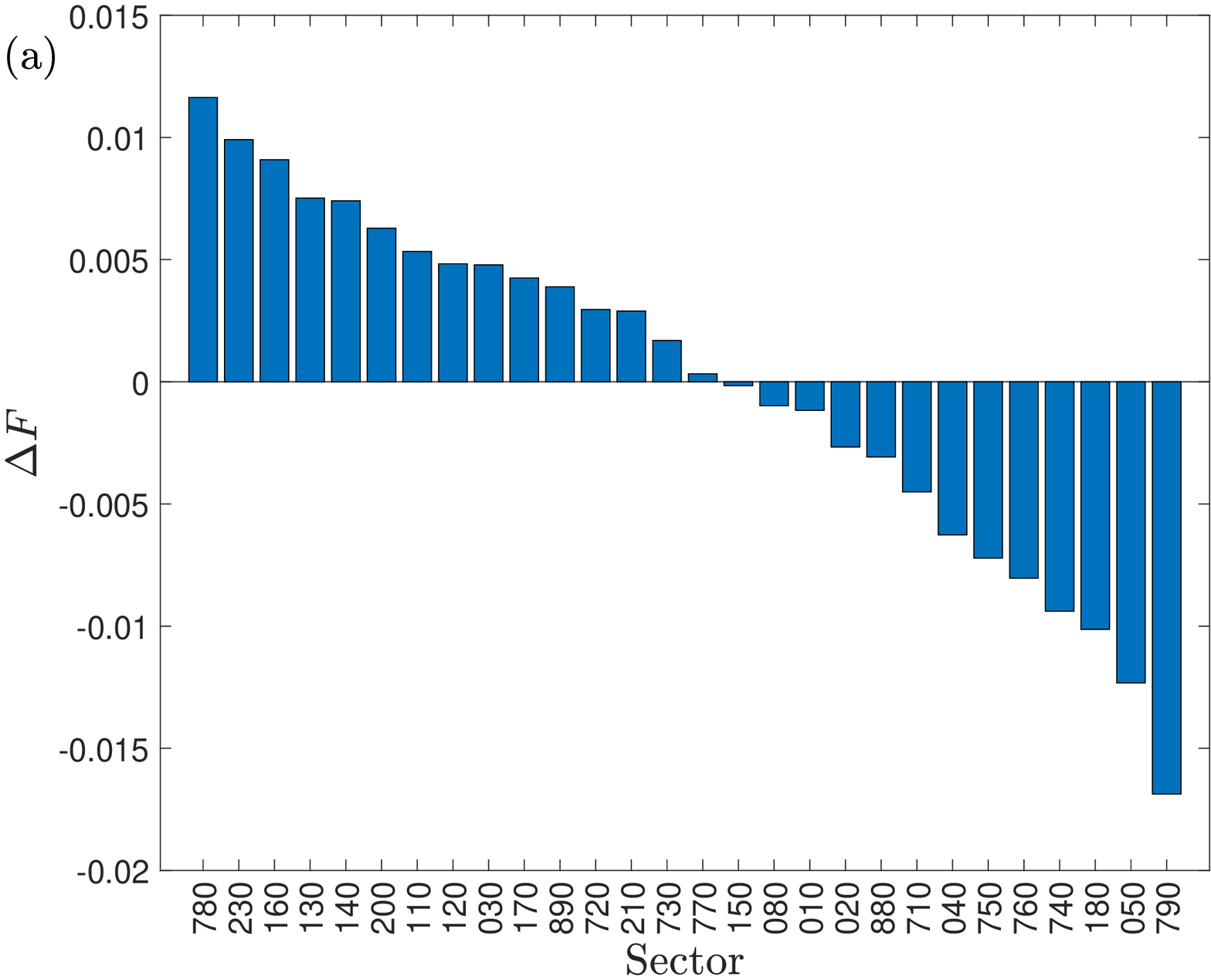} \hskip 0.2cm
  \includegraphics[width=7cm]{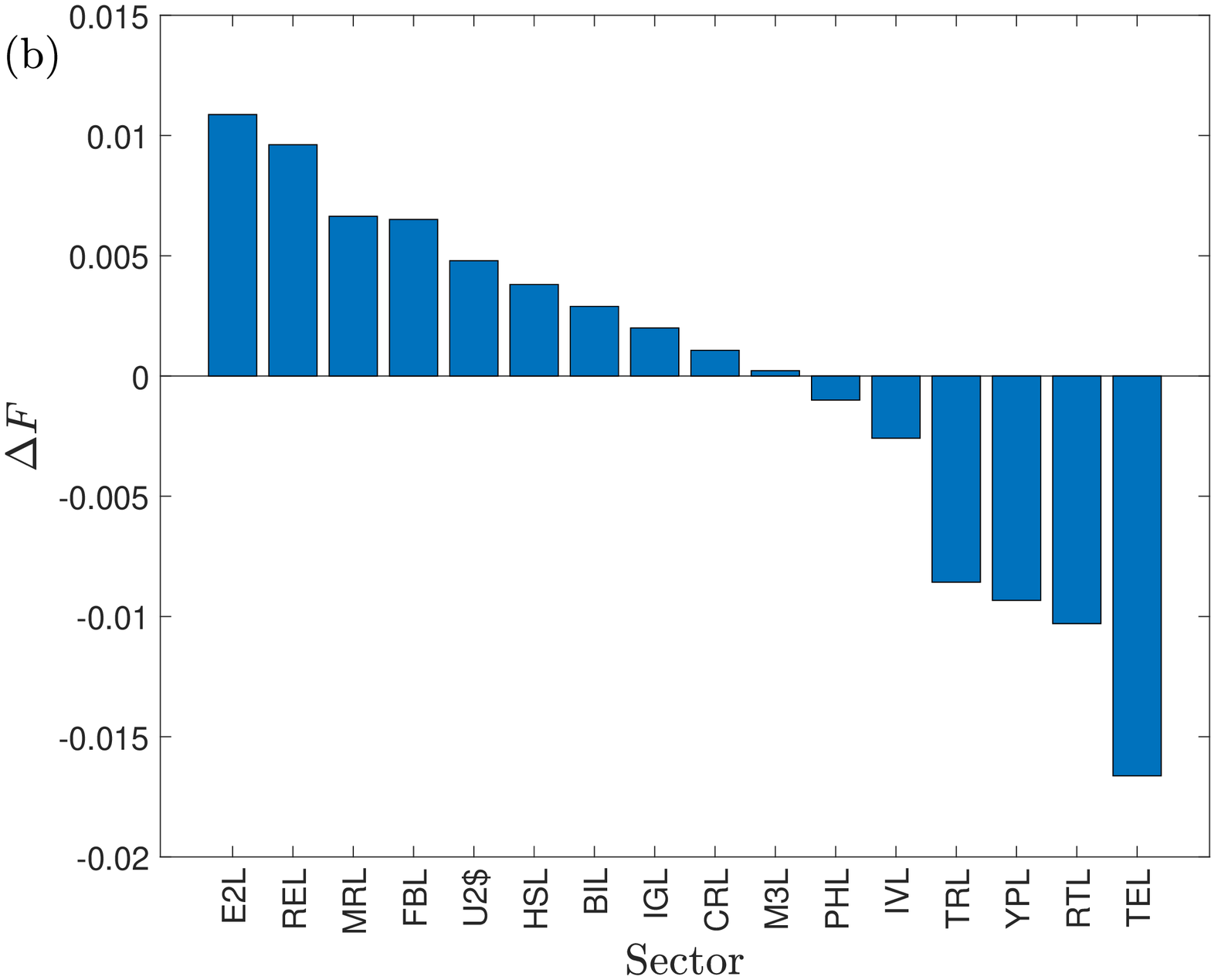}
  \caption{Degree of asymmetric information flow $\Delta{F}$ in descending order of industry sectors. (a) Chinese stock market sectors. (b) USA stock market sectors.}
  \label{Fig:STE:Sector:dTs:sorted}
\end{figure}

We illustrate in Figure~\ref{Fig:STE:Sector:dTs:sorted} the average degree of asymmetric information flow ${\Delta{F}}$ of the sectors in descending order for the two stock markets.

Among all the 28 Chinese sectors, the {\textit{bank}} sector (code 780) has the highest ${\Delta{F}}$ value, while the ${\Delta{F}}$ value of the {\textit{non-bank financial}} sector is the lowest. This finding suggests that the {\textit{bank}} sector has the highest net outflow of information and is thus the most influential sector, while the {\textit{non-bank financial}} sector is the most influenced sector. If we regard the Chinese stock market as an information transfer system, the {\textit{bank}} sector is a big information source, influencing other sectors, while the {\textit{non-bank financial}} sector is a big information sink, influenced by other sectors. Concerning the absolute ${\Delta{F}}$ value, we find that the {\textit{biotechnology}} (code 150) is the closest one to zero, which indicates that the strength of information outflows and inflows are approximately equal and there is little net information transferred between the {\textit{biotechnology}} sector and the whole market.

Among all the 16 US sectors, the {\textit{energy}} sector (code E2L) has the highest ${\Delta{F}}$ value, while the ${\Delta{F}}$ value of the {\textit{technology}} sector (code TEL) is the lowest. This suggests that the {\textit{energy}} sector has the highest net outflow of information and is thus the most influential sector, while the {\textit{technology}} sector is the most influenced sector. Therefore, the {\textit{energy}} sector is a big information source, influencing other sectors, while the {\textit{technology}} sector is a big information sink, influenced by other sectors. When we consider the absolute ${\Delta{F}}$ value, we find that the {\textit{appliances}} sector (code M3L) is the closest one to zero, which indicates that the strength of information outflows and inflows are approximately equal and there is little net information transferred between the {\textit{appliances}} sector and the whole market.

Although the sectors in both markets are similar, they play different roles in the two information transfer processes. For instance, the {\textit{real estate}} sector is an information sink in the Chinese market but an information source in the US market. These results highlight the importance of the {\textit{real estate}} sector in driving economic output in China and its less significant role in the US.


\subsection{Yearly Evolution of Symbolic Transfer Entropy and Degree of Asymmetric Information Flow}

\begin{figure}[h]
  \centering
  \includegraphics[width=11cm,height=10cm]{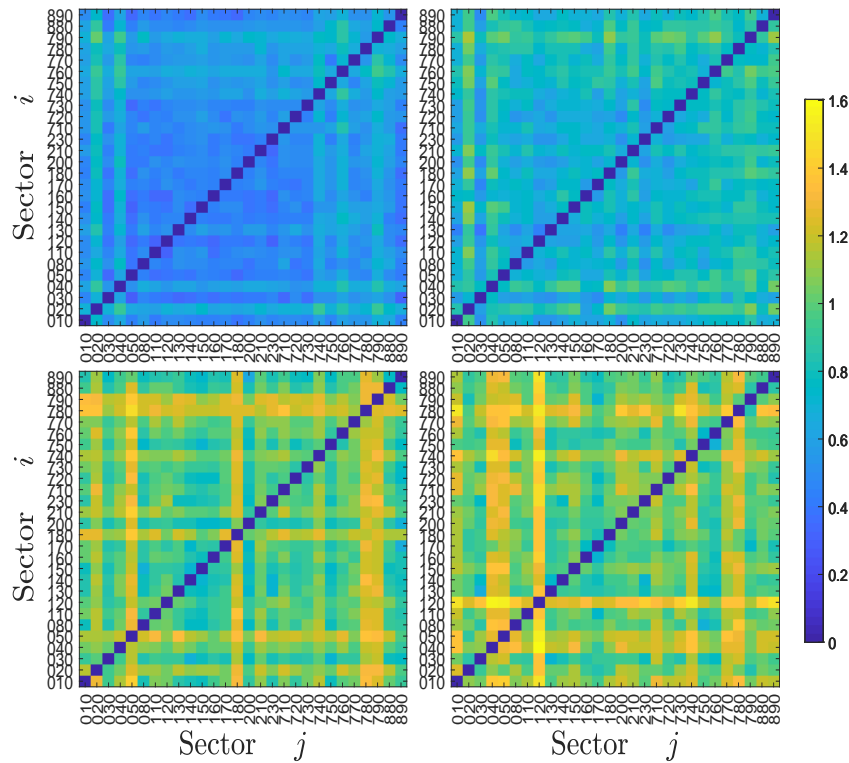}\
  \caption{Heat maps of the symbolic transfer entropy matrix $T^S_{i,j}$ for four years (2000, 2003, 2007, and 2011) of the Chinese stock market.}
  \label{Fig:STE:Sector:TS:CN:4y}
\end{figure}

\begin{figure}[h]
  \centering
  \includegraphics[width=11cm,height=10cm]{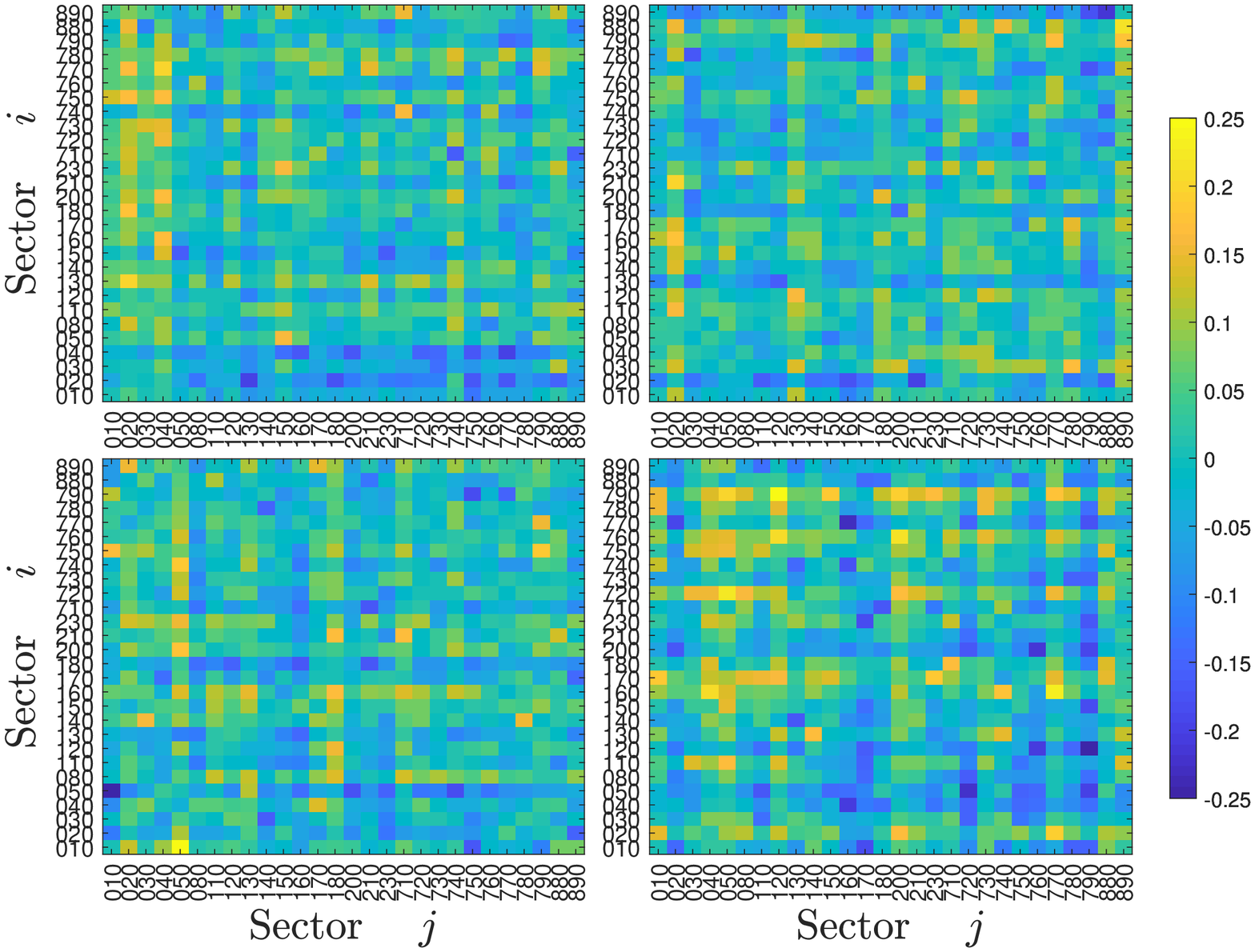}
  \caption{Heat maps of the symbolic transfer entropy matrix $\Delta{T}^S_{i,j}$ for four years (2000, 2003, 2007, and 2011) of the Chinese stock market.}
  \label{Fig:STE:Sector:dTS:CN:4y}
\end{figure}

\begin{figure}[h]
  \centering
  \includegraphics[width=11cm,height=10cm]{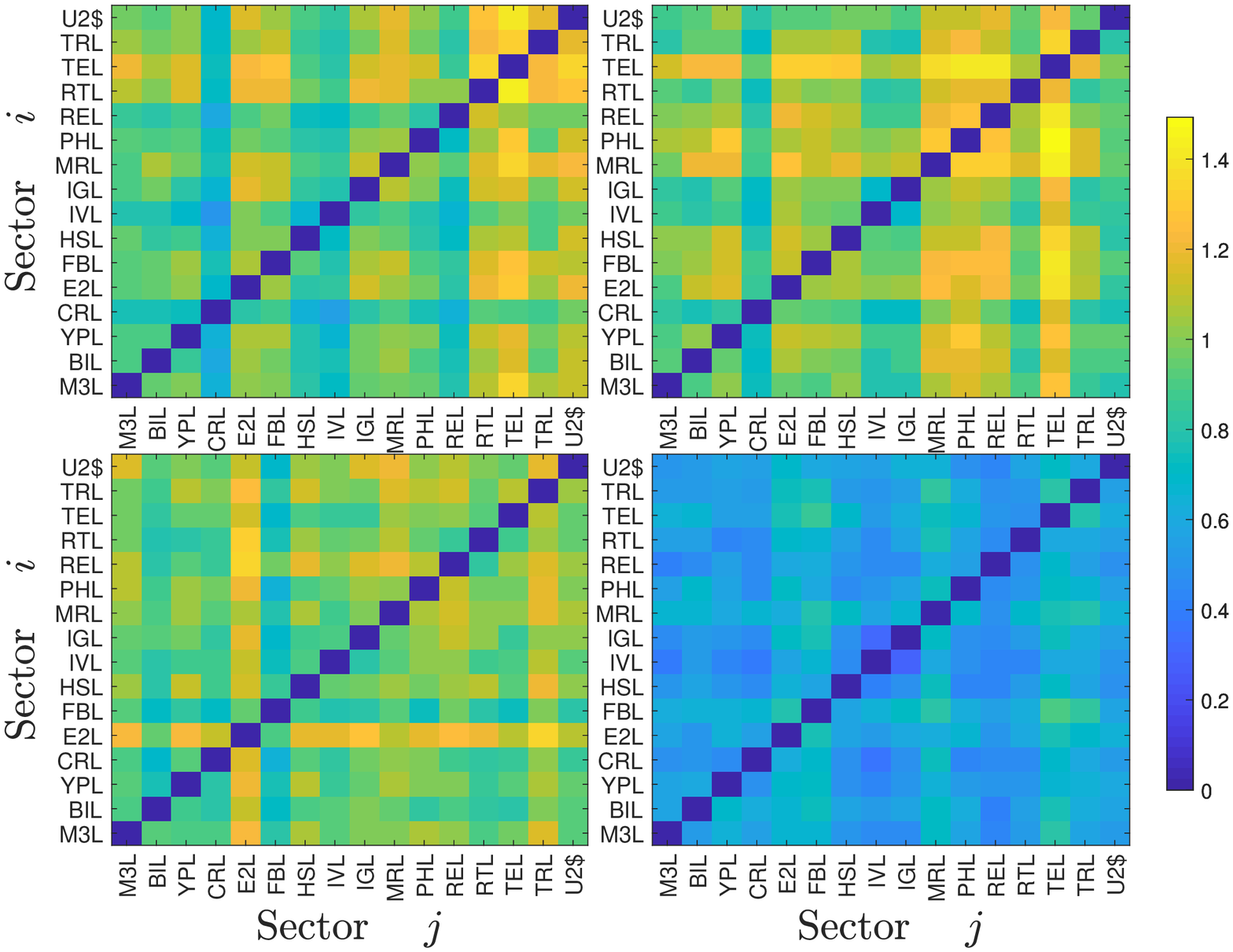}
  \caption{Heat maps of the symbolic transfer entropy matrix
  $T^S_{i,j}$ for four years (2000, 2003, 2007, and 2011) of the USA stock market.}
  \label{Fig:STE:Sector:TS:US:4y}
\end{figure}

\begin{figure}[h]
  \centering
  \includegraphics[width=11cm,height=10cm]{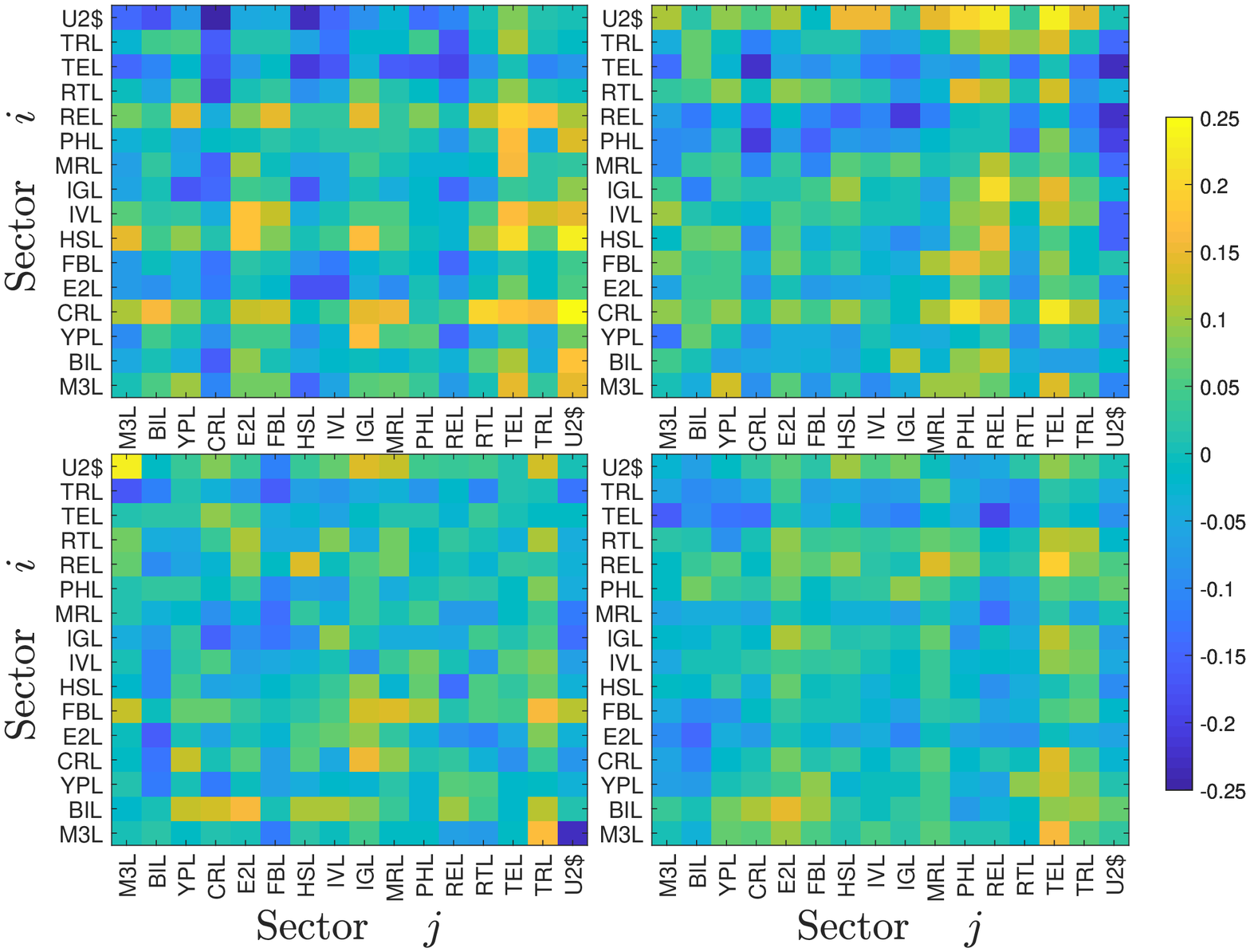}
  \caption{Heat maps of the symbolic transfer entropy matrix $\Delta{T}^S_{i,j}$ for four years (2000, 2003, 2007, and 2011) of the USA stock market.}
  \label{Fig:STE:Sector:dTS:US:4y}
\end{figure}

Economic sectoral relationships are known to be unstable and change over time. For example, Bernanke (2016) highlighted the changing correlation between the energy and industrial sectors in the US over the last decade. To qualify the evolution of information flows over time, we calculated the symbolic transfer entropy matrix $T^S(t)$ and the asymmetric average information flow $\Delta{T}^S(t)$ for each year $t$. The four $T^S(t)$ heat maps of the Chinese stock market for years 2000, 2003, 2007, and 2011 are illustrated in {{Figure~\ref{Fig:STE:Sector:TS:CN:4y} }}, and the four $T^S(t)$ heat maps of the US stock market for years 2000, 2003, 2007, and 2011 are
illustrated in {{Figure~\ref{Fig:STE:Sector:TS:US:4y} }}, respectively. For the Chinese stock market, it is found that the heat maps share some pattern of similarity. For instance, some relative bright lines emerge vertically and horizontally, echoing the pattern in {{Figure~\ref{Fig:STE:Sector:TS:CN:4y} }}. However, these heat maps also exhibit remarkable differences. The most significant feature is that the heat maps become brighter over time, indicating that there are more information transfers among different sectors with the development of the stock market. The corresponding four heat maps of the asymmetric information flow $\Delta{T}^S(t)$ are shown in {{Figure~\ref{Fig:STE:Sector:dTS:CN:4y} }}. A similar evolution of patterns is observed in the US stock markets, which is shown in {{Figure~\ref{Fig:STE:Sector:TS:US:4y} }} and {{Figure~\ref{Fig:STE:Sector:dTS:US:4y} }}. However, we do not observe a monotonic increase in information flows among the US sectors, in which the information flows among sectors were smaller in 2011.

To further quantify the evolution of information flows, we calculated the average of the symbolic transfer entropy matrix $T^S(t)$ for each year $t$ as follows:
\begin{subequations}
\begin{equation}
   \left\langle{T}^S(t)\right\rangle=\frac{1}{n(n-1)}\sum_{i\neq{j}}{{T}_{i\to{j}}^S(t)},
\label{Eq:Ave:Ts:t}
\end{equation}
where the diagonal with $i=j$ is not included, and the average asymmetric information flow $\Delta{T}^S(t)$ for each year $t$ is measured as follows:
\begin{equation}
   \left\langle\Delta{T}^S(t)\right\rangle=\frac{2}{n(n-1)}\sum_{i=1}^n\sum_{j=1}^i\big|\Delta{T}_{i\to{j}}^S(t)\big|,
\label{Eq:Ave:dTs:t}
\end{equation}
\label{Eq:Ave:Ts:dTs:t}
\end{subequations}
where the lower triangle (i.e., the part with $i\leq{j}$) is not included. {{We note that there are no objective criteria to determine the window size. Too long windows will result in too few data points and vague evolution paths, while too short windows lead to less statistics and more noise \cite{Song-Tumminello-Zhou-Mantegna-2011-PRE}. The choice of one year is a trade-off.}}


\begin{figure}[h]
  \centering
  \includegraphics[width=7cm]{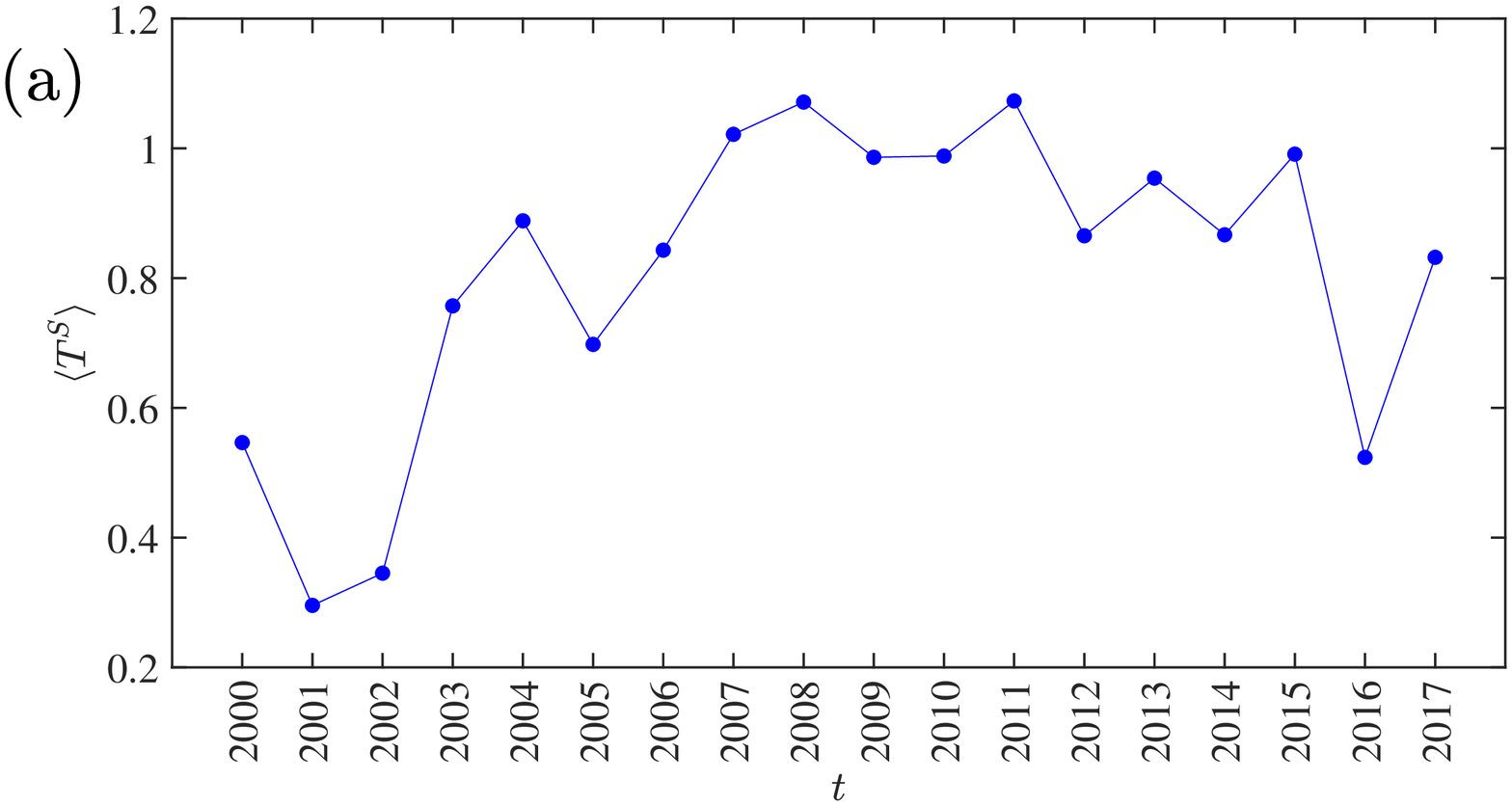}\hskip 0.2cm
  \includegraphics[width=7cm]{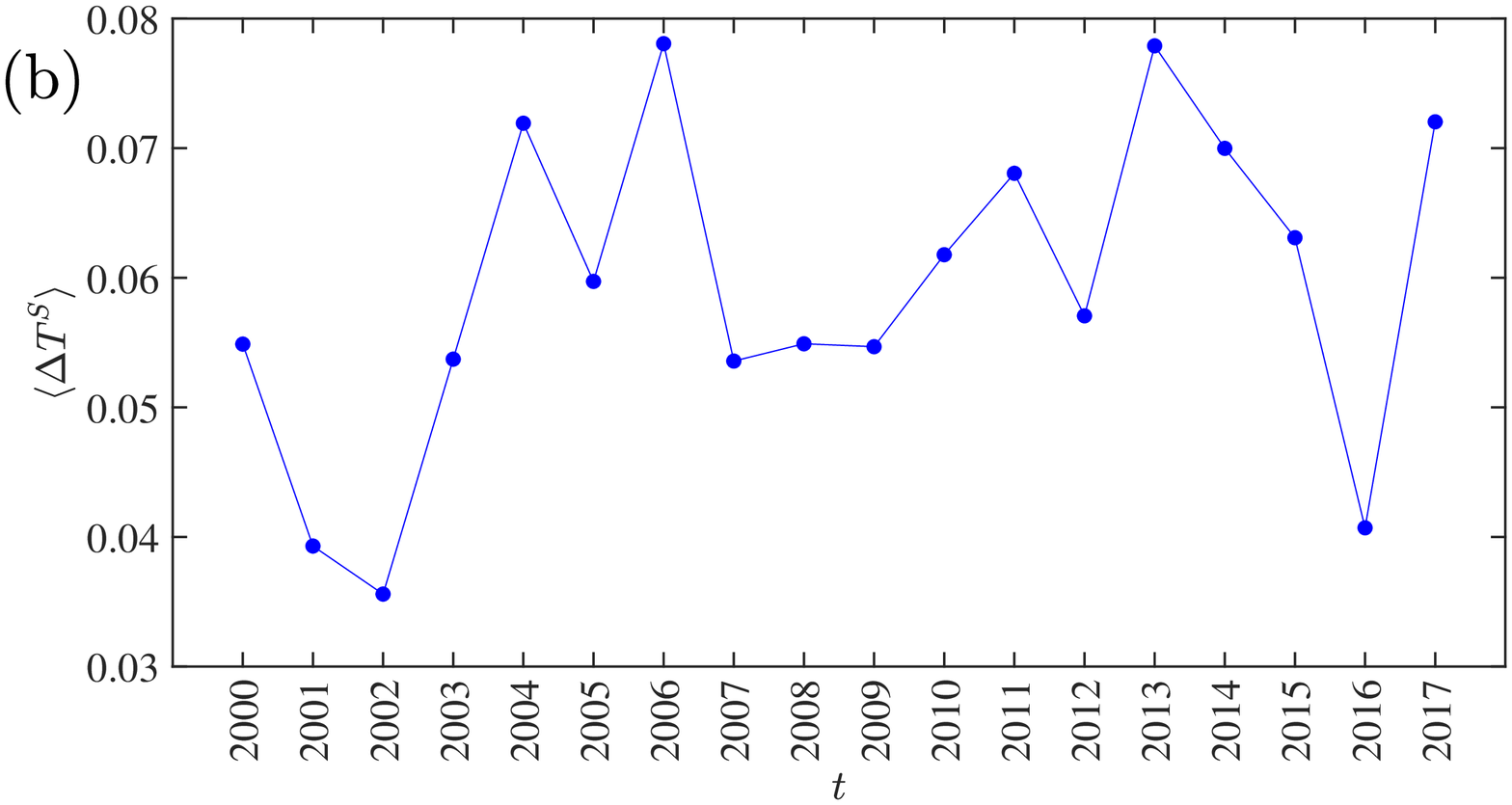}
  \includegraphics[width=7cm]{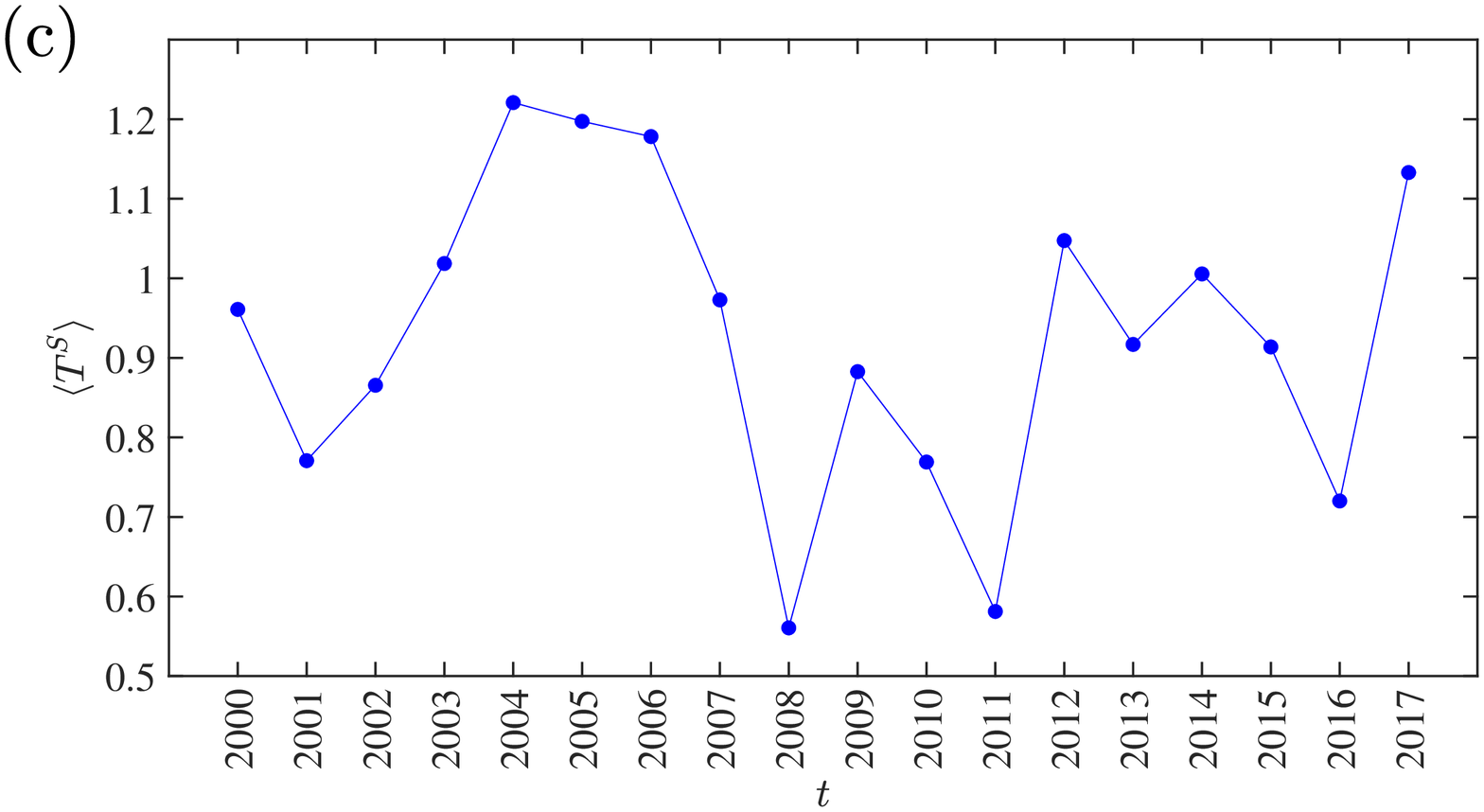}\hskip 0.2cm
  \includegraphics[width=7cm]{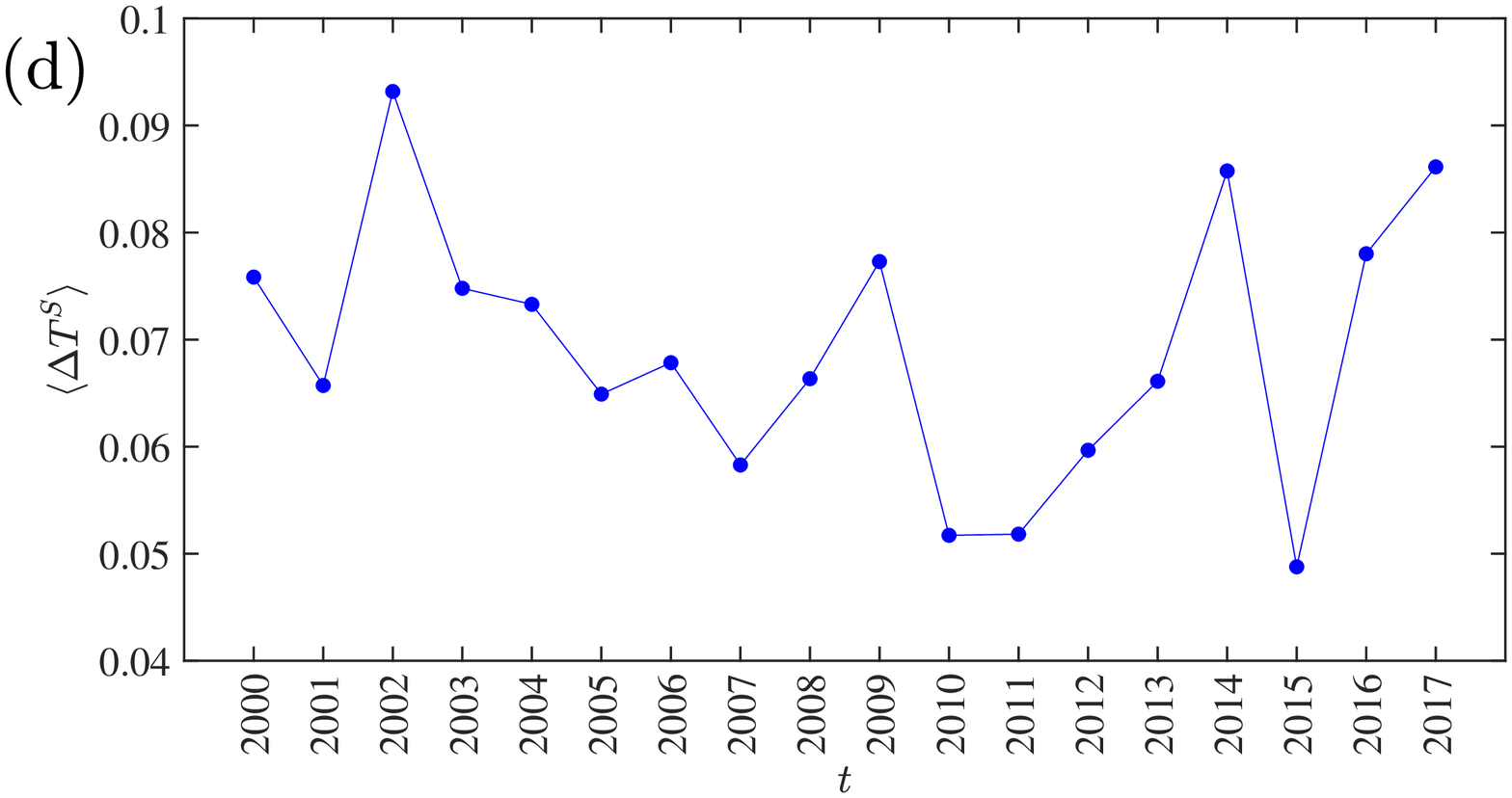}
  \caption{({\bf a}) Evolution of the average symbolic transfer entropy $\langle{T^S(t)}\rangle$ from 2000 to 2017 of the Chinese stock market. ({\bf b}) Evolution of the average
  asymmetric information flow $\langle{\Delta{T}^S(t)}\rangle$ from 2000 to 2017 of the Chinese stock market. ({\bf c}) Evolution of the average symbolic transfer entropy
  $\langle{T^S(t)}\rangle$  from 2000 to 2017 of the USA stock market. ({\bf d}) Evolution of the average asymmetric information flow $\langle{\Delta{T}^S(t)}\rangle$ from 2000 to
  2017 of the USA stock market.}
  \label{Fig:STE:Sector:TSave:t}
\end{figure}

The evolutionary trajectories of the average symbolic transfer entropy $\langle{T^S(t)}\rangle$  and the average asymmetric information flow $\langle{\Delta{T}^S(t)}\rangle$ from 2000 to 2017 of the Chinese stock market are presented in Figure~\ref{Fig:STE:Sector:TSave:t}a and Figure~\ref{Fig:STE:Sector:TSave:t}b, respectively, while the results for the US stock market are presented in Figure~\ref{Fig:STE:Sector:TSave:t}c and Figure~\ref{Fig:STE:Sector:TSave:t}d. For the Chinese stock market, we observe two local minima around 2001 and 2016 for $\langle{T^S(t)}\rangle$ and three local minima around 2001, 2008, and 2016 for $\langle{\Delta{T}^S(t)}\rangle$. This observation is of particular interest, because the three periods correspond to key periods of market volatility associated with the market crashes in June 2001 \citep{Zhou-Sornette-2004a-PA}, December 2007 \citep{Jiang-Zhou-Sornette-Woodard-Bastiaensen-Cauwels-2010-JEBO}, June 2009 \citep{Jiang-Zhou-Sornette-Woodard-Bastiaensen-Cauwels-2010-JEBO}, June 2015 \citep{Sornette-Demos-Zhang-Cauwels-Filimonov-Zhang-2015-JIS}, and January 2006 \citep{Wei-Zhang-Xiong-2017-cnJMSC}. For the US stock market, we observe four local minima around 2001, 2008, 2011, and 2016 for $\langle{T^S(t)}\rangle$ and three local minima around 2001, 2011, and 2015 for $\langle{\Delta{T}^S(t)}\rangle$, which correspond to the 9/11 terrorist attack in 2001 \citep{Charles-Darne-2006-EconM}, the subprime mortgage crisis in 2007 \citep{Demyanyk-Hemert-2011-RFS}, the July--August 2011 stock market crash \citep{Jayech-2016-EJOR}, and the 2015--16 stock market selloff beginning in the United States on 18 August 2015. It is documented for other types of networks that the structure of networks usually changes around large market movements (see~\cite{Han-Xie-Xiong-Zhang-Zhou-2017-FNL} and the references therein).

We conclude that, during market turmoil periods, both the average information transfer and the average asymmetric information flow are lower than in stable states. This conclusion is not surprising. During bubbles and antibubbles, investors exhibit stronger convergence in decision making. The majority of investors buy stocks during bubbles and sell stocks during antibubbles. Although stock markets have higher volatility during periods of turmoil, investors' actions are more synchronized. In other words, stock markets are more integrated during periods of turmoil than during  periods of stability.

\section{Conclusions}
\label{S1:conlude}

In this work, we compared the information transfer between industry sectors in the Chinese and US stock markets based on their symbolic transfer entropy. We used daily returns of key sector indices from 2000 to 2017. The results of this work offer several important insights into information flows between industry sectors. First, we find that the most active sector in information exchange is the {\textit{non-bank financial}} sector in the Chinese market and the {\textit{technology}} sector in the US market. Second, concerning the net information flow of individual sectors, we find that the main information source is the {\textit{bank}} sector in the Chinese market and the {\textit{energy}} sector in the US market, while the information sink is the {\textit{non-bank financial}} in the Chinese market and the {\textit{technology}} sector in the US market. The two information sinks with the largest net information inflow in the two markets are exactly the two most active sectors with the largest information transfer. Third, the same sector may play different roles in the two markets. For example, the {\textit{real estate}} sector is an information sink in the Chinese market but an information source in the US market. Thus, the US stock {{market is}} expected to react to demand related {{to}} news originating from the housing sector, such as building approvals, whereas in China this is not the case since the markets are driven by supply side factors such {{as}} changes in bank lending.

We also investigated the evolution of the yearly information transfer for both markets. It is found that the local minima of the average symbolic transfer entropy $\langle{T^S(t)}\rangle$ and the average asymmetric information flow $\langle{\Delta{T}^S(t)}\rangle$ correspond to periods of market turmoil. We argue that stock markets are more integrated during periods of turmoil than in stable periods, which results in smaller entropy.

Note that while there have been several studies that use entropy-based techniques to predict market fluctuations and crashes  \citep{Maasoumi-Racine-2002-JEm,Eom-Oh-Jung-2008-PA,Lahmiri-2014-FNL,Zhou-Zhan-Cai-Tong-2015-Entropy,Zou-Yu-He-2015-Entropy,Benedetto-Giunta-Mastroeni-2016-EE} or measures \citep{Hou-Liu-Gao-Cheng-Song-2017-Entropy,Gu-2017-PA,Begusic-Kostanjcar-Kovac-Stanley-Podobnik-2018-Complexity}, {{in}} this study we argue that the average symbolic transfer entropy $\langle{T^S(t)}\rangle$ and the average asymmetric information flow $\langle{\Delta{T}^S(t)}\rangle$ do not have a direct predictive power for market crashes. Further research is required to better understand the dynamics of market crashes, which are likely not driven by historical correlations but rather by behavioral factors.

\section*{Acknowledgements}

{This work} was partly supported by the National Natural Science Foundation of China (U1811462, 71532009), the Shanghai Philosophy and Social Science Fund Project (2017BJB006), the Program of Shanghai Young Top-notch Talent (2018), and the Fundamental Research Funds for the Central Universities.


\end{document}